\documentclass[9pt,twocolumn,twoside]{osajnl}

\journal{ao} 

\setboolean{shortarticle}{false}

\usepackage{color} 
\usepackage[normalem]{ulem} 

\ifthenelse{\boolean{shortarticle}}{\colorlet{color2}{color2b}}{\colorlet{color2}{color2}} 

\title{Analytical model for ring heater thermal
compensation in the Advanced Laser
Interferometer Gravitational-wave Observatory}

\author[1,*]{Joshua Ramette}
\author[2]{Marie Kasprzack}
\author[3]{Aidan Brooks}
\author[4]{Carl Blair}
\author[5]{Haoyu Wang}
\author[3]{Matthew Heintze}

\affil[1]{Hillsdale College, Hillsdale, Michigan 49242, USA}
\affil[2]{Louisiana State University, Baton Rouge, Louisiana 70803, USA}
\affil[3]{California Institute of Technology, Pasadena, California 91125, USA}
\affil[4]{University of Western Australia, Crawley WA 6009, Australia}
\affil[5]{University of Birmingham, Edgbaston, Birmingham, West Midlands B15 2TT, UK}

\affil[*]{Corresponding author: jramette@hillsdale.edu}

\dates{Received 30 December 2015; revised 23 February 2016; accepted 24 February 2016; posted 25 February 2016 (Doc. ID 256494);
published 28 March 2016}

\ociscodes{(220.1000) Aberration compensation; (350.6830) Thermal lensing; (120.6780) Temperature; (120.2230) Fabry-Perot; (120.3180) Interferometry.}

\doi{\url{http://dx.doi.org/10.1364/AO.55.002619}}

\begin{abstract}
Advanced laser interferometer gravitational-wave detectors use high laser power to achieve design sensitivity. A
small part of this power is absorbed in the interferometer cavity mirrors where it creates thermal lenses, causing
aberrations in the main laser beam that must be minimized by the actuation of “ring heaters,” which are additional
heater elements that are aimed to reduce the temperature gradients in the mirrors. In this article we derive
the first, to the best of our knowledge, analytical model of the temperature field generated by an ideal ring heater.
We express the resulting optical aberration contribution to the main laser beam in this axisymmetric case. Used in
conjunction with wavefront measurements, our model provides a more complete understanding of the thermal
state of the cavity mirrors and will allow a more efficient use of the ring heaters in the Advanced Laser
Interferometer Gravitational-wave Observatory.

\end{abstract}

\setboolean{displaycopyright}{true}

\begin{document}

\maketitle
\thispagestyle{fancy}

\ifthenelse{\boolean{shortarticle}}{\ifthenelse{\boolean{singlecolumn}}{\abscontentformatted}{\abscontent}}{}
\section{ Introduction}
Laser interferometer gravitational-wave detectors use kilometer-scale Fabry-Perot Michelson interferometers to search for astrophysical gravitational-wave signals \cite{Cadonati2015}. To maximize their sensitivity, advanced gravitational-wave detectors use continuous high power lasers, and will eventually store up to 750 kW, when operating at their maximum power, in the Fabry-Perot cavities that constitute the interferometer arms \cite{ligo2015,virgo2015}.\\
Absorption of around 0.4 W of laser power \cite{ligo2015} in the cavity mirrors, called test masses, generates a temperature gradient of a few Kelvin across each test mass \cite{HV1990}, creating a thermal lens \cite{Rocchi2012}. This thermal lens, even at lower power, produces optical aberrations in the main laser beam, compromising the sensitivity of the detectors \cite{HV1993,Strain1994,Winkler1991,VirgoSR2}. Uncompensated scattering of the gravitational waves (GW) audio sidebands from thermal lenses in the signal recycling cavity shifts the frequency response of the interferometer away from the operating point. Additionally, scattering (or losses) of the GW signal anywhere in the path between the arms and the detection photodiodes represents a direct loss for any squeezed vacuum injected into that path, reducing, in some cases dramatically, the efficiency of the reduction of shot noise. Hello and Vinet have developed analytical models of the thermal lens caused by absorption of power from the main laser beam in the test masses as well as models of the resulting aberrations \cite{HV1990, HV1990_2, Vinet2009}.\\
Several thermal compensation systems \cite{Rocchi2012,Lawrence2003} are placed in the detectors in order to mitigate these aberrations. A component called a "ring heater" \cite{Lawrence2004,Luck2004} heats the outer edge of each test mass to reduce the temperature gradient in the test masses. To apply the appropriate correction we must know how the ring heater contributes to the temperature field in each test mass \cite{Brooks2013}. The design and effect of a ring heater on a test mass has been studied with finite element analysis tools \cite{Lawrence2003, Degallaix2006}, or partially studied analytically \cite{Arain2010}. Besides providing a more intuitive understanding of the compensation mechanism,  analytical models of the thermal lens from both absorbed ring heater and laser power will allow us to apply the proper ring heater power at any time via a state estimation algorithm, leading to minimize the total thermal lens and the aberration caused by the main laser beam, and contributing to improve and stabilize the sensitivity of the detectors.\\
Here, following the methodology of \cite{HV1990}, we derive an analytical model of the temperature field caused by absorption of power from an ideal ring heater and express the resulting thermal lens for a general test mass. From this temperature field we then use the results of Hello and Vinet to calculate the optical aberration due to the temperature dependent index of refraction of the fused silica test masses. By comparing the model to in situ measurements of the thermal lens, we then estimate the effective absorption of ring heater power experienced by the test masses in Advanced LIGO (Laser Interferometer Gravitational-wave Observatory).
\\
\section{Analytical Ring Heater Temperature Field Derivation}

\begin{figure}[!h]
\centerline{\includegraphics[width= 0.5\textwidth]{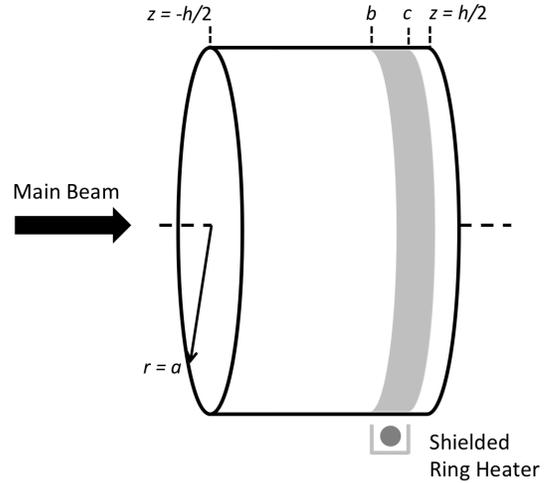}}
\caption{Diagram of cylindrical test mass. Main laser beam is incident from the left upon the surface of the mirror. Absorption of power from shielded ring heater modeled as constant on the region $b \le z \le c$ is represented by the gray area on the barrel.}
\label{TMschematic}
\end{figure}

The test mass is modeled as an axially symmetric cylinder of fused silica, with radius $a$ and depth $h$ (Fig. \ref{TMschematic}). The temperature field evolution in the test mass caused by absorption of power at the surface is a solution of the Fourier heat equation:
\begin{equation}
\rho C \frac{\partial T}{\partial t}
- K  \nabla^2T=0
\label{heat equation}
\end{equation}
where $\rho =$ mass density, $C = $ specific heat, and $K =$ thermal conductivity and we choose cylindrical coordinates for $T(t,r,z)$ such that $0\le r \le a$ and $-h/2 \le z \le h/2$.
\begin{table}[h]
\caption{Parameter values of Advanced LIGO fused silica test masses used for numerical results of analytical model} 
\centering
\begin{tabular}{c rrrrrrr}
\hline 
Symbol & Value & Description \\
\hline
$h$ & 0.2 m & Depth of test mass cylinder \\
$a$ & 0.17 m & Radius of test mass cylinder \\
$\rho$ &  2202 kg m$^{-3}$ & Mass density \\
$C$ & 772 J kg$^{-1}$ & Specific heat \\
$K$ & 1.38 W m$^{-1}$ & Thermal conductivity \\
$\sigma$' & 0.9 $\times$ 5.67 $\times$ 10$^{-8}$ &  Emissivity$\times$Stefan-Boltzmann \\
$T_{ext}$ & 300 K & Ambient temperature \\
$\gamma P$ & 1 W & Ring heater absorbed power \\
$b$ & 0.057 m & Ring heater power boundary \\
$c$ & 0.076 m & Ring heater power boundary \\
$dn/dT$ & 0.86 $\times$ 10$^{-5}$ K$^{-1}$ & Refractive index $T$ dependence\\
$\tau$ & $4\sigma'T_{ext}^3 a/K$ = 0.679 & Reduced radiation constant\\
\hline
\end{tabular}
\label{parameters}
\end{table}

The ring heater is a glass torus wrapped in nichrome wire positioned around the barrel of the test mass near one end. Electrical power is dissipated in the nichrome wire, heating the glass torus, which radiates power onto the test mass in the infrared region that is absorbed efficiently by the silica test mass surface. We model the ring heater power as incident on the test mass surface at $r = a$ in a thin band of constant power around the circumference of the cylinder extending from $z = b$ to $z = c$. The intensity radiated from the ring heater $I_{RH}$ impinging in this region of the surface of the test mass is the power $P$ emitted by the ring heater divided by the surface area of the constant power band $2 \pi a (c-b)$:
\begin{equation}
I_{RH}(z) = \frac{1}{2 \pi a (c-b)}
\begin{cases}
P, & \text{ $b \le z \le c$} \\
0, & \text{elsewhere}
\end{cases}
\label{PRH}
\end{equation}
This is a good approximation for the absorbed ring heater power since the ring heater radiates close to the test mass surface and is equipped with a shield to reflect and concentrate the absorbed power into a thin region.
To determine boundary conditions for each surface of the test mass, we equate the power radiated with the power conducted through the silica. Since the test mass is suspended from thin silica wires in ultra-high vacuum, we approximate the model heat losses through radiation only. We linearize the net radiated power $F$ since the temperature of the heated test mass is just a few Kelvin above the ambient temperature $T_{ext} \approx 300$ K as in \cite{HV1990}:
\begin{equation}
F = 4 \sigma'T_{ext}^3 (T - T_{ext})
\end{equation}
where $\sigma'$ is the Stefan-Boltzmann constant times the emissivity. This linearization means that the temperature field scales linearly with the applied ring heater power, which allows us to construct the temperature field in the test mass as the linear superposition of the contributions from the ring heater and the main laser beam separately.\\
Due to the low thermal conductivity of fused silica, it takes around 30 hours for the test mass to reach steady state, requiring us to model the time dependence of the temperature field in order to efficiently apply a ring heater correction. We first derive the steady state solution and then use this solution to derive the transient solution.
\subsection{Steady state solution}
To derive the steady state solution, we consider the temperature field to be the sum of the ambient temperature $T_{ext}$ and a steady state solution $T_{ss}(r,z)$:
\begin{equation}
T(r,z) = T_{ext} + T_{ss}(r,z)
\end{equation} 
Then for the steady state temperature field we must solve the Laplacian:
\begin{equation}
\nabla^2 T_{ss}(r,z)=0
\end{equation}

Since the ring heater only delivers power to the radial side of the test mass we have the following boundary conditions:
\begin{align}
&-\frac{\partial T_{ss}}{\partial r} (a,z) = \frac{\tau}{a} T_{ss}(a,z) - \frac{\gamma}{K}  I_{RH}(z)
\label{b1} \\
&-\frac{\partial T_{ss}}{\partial z} (r,-h/2) = - \frac{\tau}{a} T_{ss}(r,-h/2)
\label{b2} \\
&-\frac{\partial T_{ss}}{\partial z} (r,h/2) = \frac{\tau}{a} T_{ss}(r,h/2)
\label{b3}
\end{align}
where $\tau = 4\sigma'T_{ext}^3 a/K$ is the reduced radiation constant and $\gamma$ is the fraction of total emitted ring heater power absorbed on the surface of the test mass.

The solution takes the form
\begin{align}
T_{ss}(r,z) = \sum\limits_{m=1}^\infty \Big[ A_m \cos (u_m z/a) I_0(u_m r/a) + \\
B_m \sin(v_m z/a) I_0 (v_m r/a) \Big]
\end{align}
where $I_0$ is the modified Bessel function of the first kind and $A_m,B_m,u_m,$ and $v_m$ are determined by the boundary conditions.

From conditions (\ref {b2}) and (\ref{b3}) we obtain $u_m$ and $v_m$ as the $m$th solutions to the following equations, which must be computed numerically:
\begin{align}
u_m =& \tau \cot(u_m h/2a) \notag  \\
v_m =& -\tau \tan(v_m h/2a)
\end{align}

Condition (\ref{b1}) gives the following, considering the $m$th term in the series:
 \begin{align}
&A_m \cos(u_m z/a) [I_1(u_m) u_m/a + I_0(u_m) \tau /a] \notag \\ 
+ &B_m \sin(v_m z/a) [I_1(v_m) v_m/a + I_0(v_m) \tau /a] = \frac{\gamma}{K} I_{RH}(z)
\label{mthterm}
\end{align}
To obtain $A_m$ and $B_m$ we use the Fourier orthogonality relations with the following normalizations:
\begin{align}
\int_{-h/2}^{h/2} \cos^2(\omega x) dx = \frac{\omega h + \sin(\omega h)}{2 \omega}\\
\int_{-h/2}^{h/2} \sin^2(\omega x) dx = \frac{\omega h - \sin(\omega h)}{2 \omega}
\label{normalization}
\end{align}
Applying the orthogonality relations to (\ref{mthterm}) yields:
\begin{align}
A_m [I_1(u_m) u_m/a + I_0(u_m) \tau /a] \int_{-h/2}^{h/2} \cos^2(u_m z/a) dz = \notag \\
\frac{\gamma}{K}  \int_{-h/2}^{h/2} \cos(u_m z/a)I_{RH} (z) dz
\end{align}
\begin{align}
B_m [I_1(v_m) v_m/a + I_0(v_m) \tau /a] \int_{-h/2}^{h/2} \cos^2(u_m z/a) dz = \notag \\
\frac{\gamma}{K}  \int_{-h/2}^{h/2} \cos(u_m z/a)I_{RH}(z) dz
\end{align}
Recalling from (\ref{PRH})  the definition of $I_{RH}(z)$, zero outside the interval $[b,c]$, and using the relation (\ref{normalization}) we find:
\begin{align}
A_m [I_1(u_m) u_m/a + I_0(u_m) \tau /a] \frac{u_m h/a + \sin(u_mh/a)}{2u_m/a} = \notag \\
\frac{\gamma}{K 2 \pi a (b-c)}  P \int_{b}^{c} \cos(u_m z/a)dz \notag \\
B_m [I_1(v_m) v_m/a + I_0(v_m) \tau /a] \frac{v_m h/a - \sin(v_mh/a)}{2v_m/a} = \notag \\
\frac{\gamma}{K 2 \pi a (b-c)}  P \int_{b}^{c} \sin(v_m z/a)dz
\end{align}
So that $A_m$ and $B_m$ become:
\begin{align}
A_m = \frac{2 \gamma P}{K 2 \pi a (c-b)} \frac{[\sin(v_m c/a) - \sin(v_m b/a)]}{[u_m h/a + \sin(u_mh/a)] [I_1(u_m) u_m/a + I_0(u_m) \tau /a]} \notag \\
B_m = \frac{2 \gamma P}{K 2 \pi a (c-b)} \frac{[-\cos(v_m c/a) + \cos(v_m b/a)]}{[v_m h/a - \sin(v_mh/a)] [I_1(v_m) v_m/a + I_0(v_m) \tau /a]}
\end{align}

\begin{figure}[!h]
\centerline{\includegraphics[width= 0.47\textwidth]
{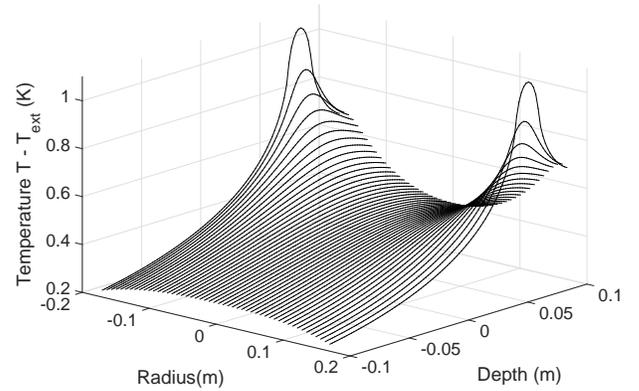}}
\caption{Steady-state temperature field in the test mass (above ambient temperature $T_{ext}$) from absorption of ring heater power in the region 0.057 m $\le z \le $ 0.076 m}
\label{sstfield}
\end{figure}

Fig. \ref{sstfield} shows the steady state temperature field solution for a test mass and ring heater with LIGO parameters. The temperature field is highest around the region of incident ring heater power and slopes off because of radiative cooling from the test mass surface.

\subsection{Transient solution}
To construct the transient solution we consider the time dependent temperature field to be the sum of the ambient temperature, the steady state solution, and a transient solution $T_{tr}$:
\begin{equation}
T(t,r,z) = T_{ext} + T_{ss}(r,z) + T_{tr}(t,r,z)
\end{equation}
The boundary conditions for $T_{tr}$ represent only radiative cooling since the transient solution disappears as $t$ approaches infinity, allowing $T(t,r,z)$ to converge to the steady state $T_{ss}$:
\begin{align}
&-\frac{\partial T_{tr}}{\partial r} (a,z) = \frac{\tau}{a} T_{tr}(a,z)
\label{bt1} \\
&-\frac{\partial T_{tr}}{\partial z} (r,-h/2) = - \frac{\tau}{a} T_{tr}(r,-h/2)
\label{bt2} \\
&-\frac{\partial T_{tr}}{\partial z} (r,h/2) = \frac{\tau}{a} T_{tr}(r,h/2)
\label{bt3}
\end{align}
In addition, we assume that at $t=0$ the temperature throughout the test mass is equal to $T_{ext}$:
\begin{equation}
T(0,r,z) = T_{ext}  \implies T_{tr}(0,r,z) = -T_{ss}(r,z)
\label{bt4}
\end{equation}
The general transient solution to the heat equation (\ref{heat equation}) is:
\begin{align}
T_{tr}(r,z) = \sum\limits_{m, p=1}^\infty \Big[ A_{mp}  e^{-\alpha_{mp} t}  \cos (u_m z/a) J_0(\zeta_p r/a) + \notag \\ B_{mp} e^{-\beta_{mp} t} \sin(v_m z/a) J_0 (\zeta_p  r/a) \Big]
\end{align}
where $J_0$ is the Bessel function of the first kind and $\alpha_{mp} = (u_m^2 + \zeta_p^2)K/(\rho a^2)$ and $\beta_{mp} = (v_m^2 + \zeta_p^2)K/(\rho a^2)$.

Conditions (\ref{bt1}) and (\ref{bt2}) give the same definitions for $u_m$ and $v_m$ as before, and condition (\ref{bt3}) gives $\zeta_p$ as the $p$th solution to the following equation:
\begin{equation}
J_1(\zeta_p)\zeta_p - \tau J_0(\zeta_p) = 0
\end{equation}
 To determine $A_{mp}$ and $B_{mp}$ we use condition (\ref{bt4}):
 \begin{align}
 \sum\limits_{m, p=1}^\infty \Big[ A_{mp}   \cos (u_m z/a) J_0(\zeta_p r/a) + B_{mp} \sin(v_m z/a) J_0 (\zeta_p  r/a) \Big] \notag \\ =  - \sum\limits_{m=1}^\infty \Big[ A_m \cos (u_m z/a) I_0(u_m r/a) + B_m \sin(v_m z/a) I_0 (v_m r/a) \Big]
 \end{align}
 or looking at the $m$th term:
 \begin{align}
\sum\limits_{p=1}^\infty \Big[ A_{mp}   \cos (u_m z/a) J_0(\zeta_p r/a) + B_{mp} \sin(v_m z/a) J_0 (\zeta_p  r/a) \Big] \notag \\ = -\Big[ A_m \cos (u_m z/a) I_0(u_m r/a) + B_m \sin(v_m z/a) I_0 (v_m r/a) \Big]
 \end{align}
 Equating and canceling the sine and cosine terms gives:
 \begin{align}
 \sum\limits_{p=1}^\infty A_{mp} J_0(\zeta_p r/a) = -A_m I_0(u_m r/a)
\label{dini1}\\
\sum\limits_{p=1}^\infty B_{mp} J_0(\zeta_p r/a) = -B_m I_0(v_m r/a)
\label{dini2}
\end{align}
The functions $J_0(\zeta_p r/a)$ form an orthogonal basis on the interval $[0,a]$ with the following orthogonality conditions:
\begin{equation}
\int_{0}^{a} J_0(\zeta_i r/a) J_0(\zeta_j r/a)rdr =  
\begin{cases}
0, & \text{ $ i \neq j $} \\
\frac{a^2}{2 \zeta_i}(\tau^2 + \zeta_i^2)J_0(\zeta_i)^2, & \text{ $i = j$ }
\end{cases}
\label{bessel_orthogonality}
\end{equation}
These conditions allow us to expand $I_0(u_m r/a)$ and $I_0(v_m r/a)$ in a Dini series:
\begin{align}
I_0(u_m r/a) = \sum\limits_{p=1}^\infty c_p^u J_0(\zeta_p r/a) \\
I_0(v_m r/a) = \sum\limits_{p=1}^\infty c_p^v J_0(\zeta_p r/a)
\end{align} 
Substituting these Dini series back into (\ref{dini1}) and (\ref{dini2}) and applying the orthogonality relations (\ref{bessel_orthogonality}) gives the coefficients $c_p^u$, $c_p^v$:
\begin{align}
 c_p^{u,v} = \frac{2 \zeta_p^2[(u_m,v_m)I_1(u_m,v_m)J_0(\zeta_p) + \zeta_p I_0(u_m,v_m)J_1(\zeta_p)]}{[(u_m,v_m)^2 + \zeta_p^2][\tau^2 + \zeta_p^2]J_0(\zeta_p)^2}
\end{align}
Using condition (\ref{bt4}) we can now express the time dependent temperature field as:
\begin{align}
T(t,r,z) =T_{ext} +  \sum\limits_{m, p=1}^\infty \Big[ A_{mp}  [e^{-\alpha_{mp} t} -1]  \cos (u_m z/a) J_0(\zeta_p r/a) + \notag \\
B_{mp} [e^{-\beta_{mp} t}-1] \sin(v_m z/a) J_0 (\zeta_p  r/a) \Big]
\notag \\ = T_{ext} +  \sum\limits_{m, p=1}^\infty \Big[ A_{m} c_p^u  [1 - e^{-\alpha_{mp} t} ]  \cos (u_m z/a) J_0(\zeta_p r/a) + \notag \\ 
B_{m}c_p^v  [1 - e^{-\beta_{mp} t}] \sin(v_m z/a) J_0 (\zeta_p  r/a) \Big] \notag \\
\label{foraberration}
\end{align}

\begin{figure}[!h]
\centerline{\includegraphics[width= .5\textwidth]{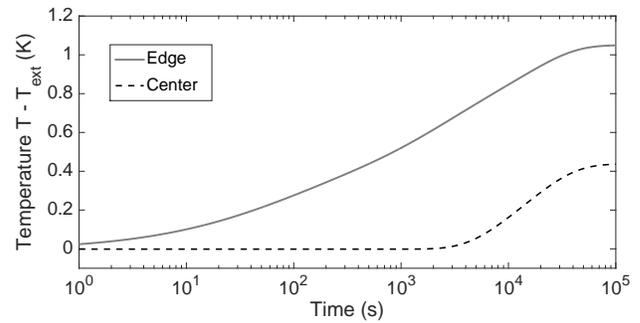}}
\caption{Temperature evolution of two points, $r=$ 0 m, $z=$ 0.07 m (dashed) and $r=$ 0.17 m, $z=$ 0.07 m (solid) for $\gamma P = 1$ W. Notice how the point $r=$ 0 m, far from the ring heater, experiences almost no heating until $t=$ 1000 s, whereas the point at $r=$ 0.17 on the surface region where the ring heater power is applied begins heating immediately.}
\label{RHtime}
\end{figure}

Fig. \ref{RHtime} shows the temperature field evolution of the transient solution for two points, near the edge and the center of the test mass. It takes around $10^5$ s (more than 27 hours) for the temperature field to reach steady state.

\section{Optical Aberration from Ring Heater Thermal Lensing}
From the temperature field, we calculate the thermal aberration experienced by a laser beam propagating through the test mass. The aberration $\psi$ is the optical path distortion (in meters) of the wavefront after propagating through the test mass relative to the wavefront in the case where the test mass has uniform temperature $T_{ext}$. The dominant effect is the thermal lens, created by the temperature dependent index of refraction. Other effects including thermo-elastic deformation and elastooptic effect will contribute to the modification of the optical path, but to a lesser extent: they represent less than 10\% of the total aberration for the fused silica \cite{Lawrence2003}. However these are relevant to the main beam through aberrations induced by reflection of the test masses and are carefully studied \cite{Lawrence2004, King2015}. In this approximation, the aberration $\psi$ due to a temperature dependent index of refraction in the test mass $dn/dT$ as a function of the radius $r$ and time $t$ is given by \cite{HV1990}:
\begin{equation}
\psi(t,r) = \frac{dn}{dT}  \int_{-h/2}^{h/2}[T(t,r,z) - T_{ext}]dz
\label{psi}
\end{equation}

Applying this equation to the analytical ring heater temperature field model (\ref{foraberration}) gives the analytical model for the aberration caused by ring heater power absorption, $\psi_{mod}(t,r)$:
\begin{equation}
\psi_{mod}(t,r) = 2\frac{dn}{dT}  \sum\limits_{m,p = 1}^{\infty}A_{mp} c_p^u \sin(u_mh/2a)(a/u_m)[1 - e^{-\alpha t}]J_0(\zeta_p r/a)
\end{equation}

\begin{figure}[!h]
\centerline{\includegraphics[width= .5\textwidth]{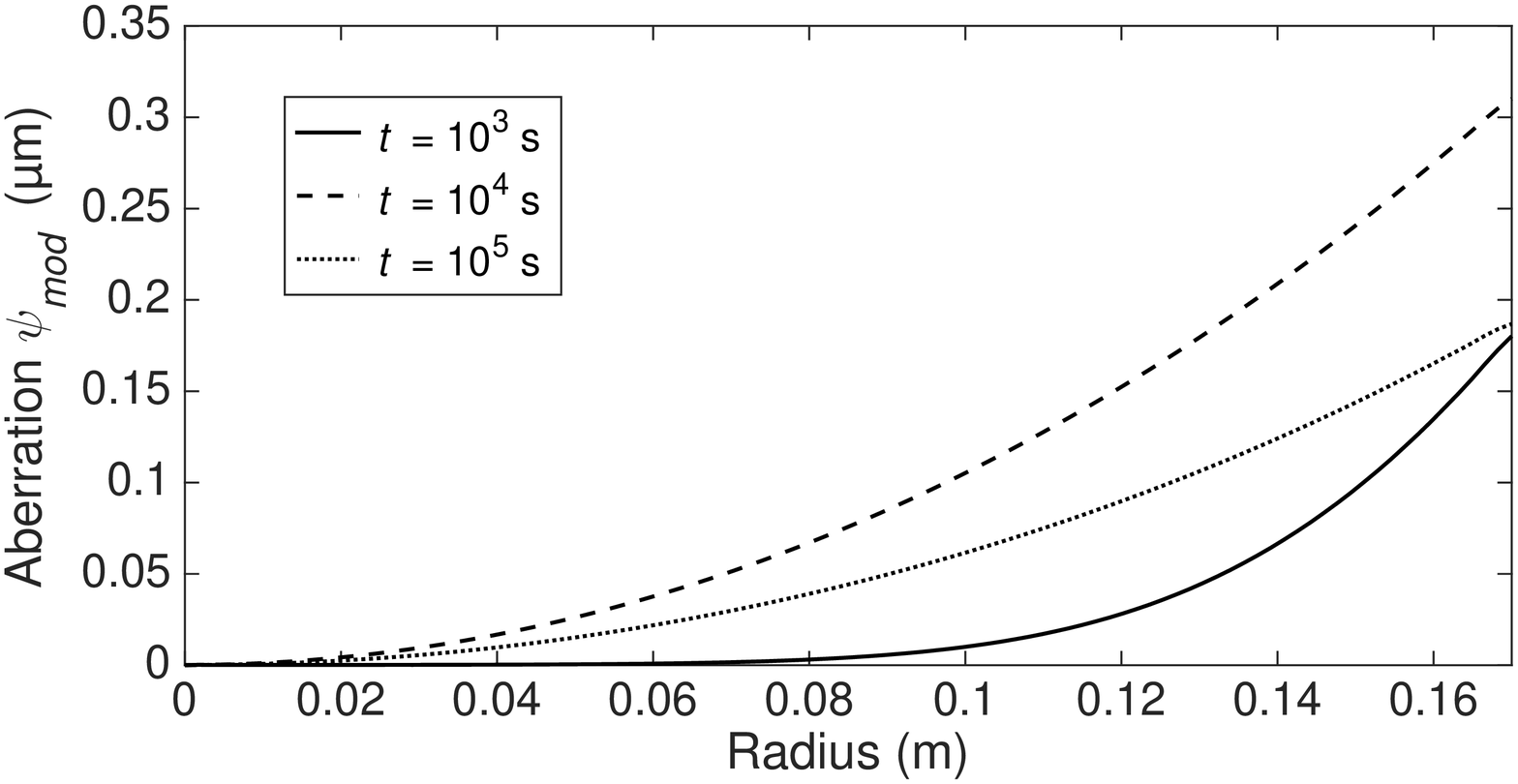}}
\caption{Thermal lens profile at $t = 10^3$ s (solid), $t = 10^4$ s (dashed), and $t = 10^5$ s (dotted). Each profile has been set to zero at $r = 0$ since only changes in $\psi$ with $r$ contribute to net aberration in the main beam. Note that the aberration first grows, especially near the edge of the test mass, and then decreases as the test mass approaches steady state due to temperture uniformization over the test mass.}
\label{aberration}
\end{figure}

Fig. \ref{aberration} shows the aberration as a function of the radial coordinate of the test mass.

\section{Hartmann Wavefront Sensor Measurements}
The Thermal Compensation System for Advanced LIGO includes Hartmann Wavefront Sensors (HWS) at each test mass \cite{Brooks2013}. These HWS measure the changes to the wavefront of an approximately 4 cm radius probe laser beam that double passes the test mass. The probe beam enters the anti-reflecting end of the test mass and reflects off of the internal side of the high-reflective coating, accumulating an aberration by passing through the bulk of the test mass twice. The probe beam has similar dimensions to the main cavity laser beam, such that its aberration is a measure of the aberration experienced by the main laser in transmission through the test mass. The measured probe beam wavefront $\psi_{meas}(t,x,y)$ is numerically reconstructed and then decomposed into a dimensionless polynomial basis in powers of $x$ and $y$ up to $n,m = 6$, where $x$ and $y$ are Cartesian coordinates:
\begin{equation}
\psi_{meas}(t,x,y) \approx \sum\limits_{n,m = 0}^{6} c_{nm}(t) x^n y^m
\end{equation}
where the coefficients $c_{nm}$ are now functions of time. \cite{HWSequations}.

The HWS computer streams relevant combinations of these coefficients to the LIGO data channels, including the ``spherical" component of the aberration, which is computed from the coefficients as follows \cite{HWSequations}:
\begin{equation}
S = c_{20} + c_{02} - \sqrt{(c_{20} - c_{02})^2 + c_{11}^2}
\label{S}
\end{equation}
$S$ is essentially the quadratic, $(S/2)r^2$, component of the wavefront; note that if the wavefront is axisymmetric, then the term under the square root is zero.

To test the analytical ring heater temperature field model we compare the model's prediction for how $S$ evolves with time to the HWS measured values for $S$. Since the model contains only axisymmetric terms, $S$ provides the principal component of the model's aberration.  To compute the model's prediction for $S$, we project the analytical model $\psi_{mod}(t,r)$ onto a polynomial basis in powers of $x$ and $y$ and use (\ref{S}) to calculate $S$ over the 4 cm radius section of the test mass that the HWS probe beam senses.

To obtain data from the HWS to test the analytical ring heater temperature field model we step the electrical power through the ring heater and capture the transient evolution of $S$. The ring heater itself takes 10-15 minutes to heat up and reach steady state after the electrical power step, meaning that the power radiated by the ring heater onto the test mass is delayed relative to the model, in which the power is modeled as a step function. However, this delay is a small fraction of the duration of the 24 hour time scale over which we compare the model and data. We measure a total of eight transients, four from each end test mass (ETM) at the ends of the X and Y arms of the interferometer (ETMX, ETMY).

The model and data allow us to estimate $\gamma$, the proportion of ring heater power absorbed in the Advanced LIGO test masses.  Indeed, all parameters of the model are determined except for $\gamma$, which scales the magnitude of the aberration. By fitting the model to the data, we calculate the fraction of ring heater power absorbed by the test mass. 
$\gamma$ is expressed by Equation (\ref{eq:gammaCalc}) as the ratio of the maximum spherical aberration between the model and the data for a given ring heater power, taking into account that the data represents a ``double pass" through the bulk of the test mass, since the HWS probe beam enters the back of the ETM and reflects off of the internal side of the reflective coating (Table \ref{gammatable}): 

\begin{equation}
\gamma = \frac{1 \mbox{ W} \times Peak\mbox{ }S_{HWS}}{2P \times Peak\mbox{ }S_{mod}}\label{eq:gammaCalc}
\end{equation}

where $P$ represents the step in the total power emitted by the ring heater, $Peak\mbox{ }S_{mod} = 2.127 \times 10^{-5}$ m is the analytical model's peak evaluated with $\gamma P = 1$ W, and the factor of 2 corrects for the double pass in the HWS measurements.

\begin{figure}[!h]
\centerline{\includegraphics[width= .5\textwidth]{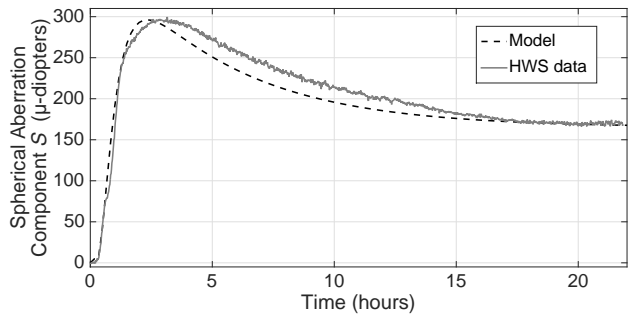}}
\caption{Comparison of HWS data (solid) and analytical model spherical aberration component (dashed) after a step in ring heater power. Model is scaled to match the data peak and $\gamma$ is estimated to be about 0.78.}
\label{heatingy1}
\end{figure}

An example of $\gamma$ estimation is presented in Fig. \ref{heatingy1}: after adjusting $\gamma$ in our model to match the data peak, the model and data show reasonable agreement for the spherical component of the aberration. Note that the model's peak lags the data slightly, suggesting that the differences between the model and data are caused by the model slightly underestimating the time constant for the temperature evolution.

\begin{table}[h]
\caption{Computation of $\gamma$. Table shows data from the four transients from each ETM. The powers steps for ETMX are smaller because a malfunction in the ETMX ring heater limited the power we could apply through the ETMX ring heater. For the model we used the same parameters as given in Table \ref{parameters}, except that $T_{ext} = 293.15$ K.} 
\centering
\begin{tabular}{c rrrrrrr}
\hline 
ETM & $Peak \mbox{ } S_{HWS}$ & Power step $P$ & $\gamma$ \\
\hline
ETMY & 1.70$\times 10^{-4}$ m & 5.89 W & 0.679 \\
ETMY & 2.36$\times 10^{-4}$ m & 6.86 W & 0.808 \\
ETMY & 2.95$\times 10^{-4}$ m & 8.87 W & 0.781 \\
ETMY & 3.15$\times 10^{-4}$ m & 9.84 W & 0.754 \\

\hline 
ETMX & 0.712$\times 10^{-4}$ m & 2.47 W & 0.678 \\
ETMX & 0.802$\times 10^{-4}$ m & 2.47 W & 0.763 \\
ETMX & 1.17$\times 10^{-4}$ m & 3.43 W & 0.800 \\
ETMX & 1.07$\times 10^{-4}$ m & 3.44 W & 0.733 \\

\hline
\end{tabular}
\label{gammatable}
\end{table}

Averaging the $\gamma$ values from each step in Table \ref{gammatable}, we estimate $\gamma$ for each test mass:\
\begin{align}
\label{gammaestimates}
\text{ETMY:   }   \gamma = 0.76 \pm 0.08 \\
\text{ETMX:   }   \gamma = 0.74 \pm 0.07 \notag
\end{align}

Because of the reflective shield surrounding the ring heater and the efficient infrared power absorption of fused silica, we expect a high absorption efficiency. The values for $\gamma$ are reasonable considering these conditions and further confirm the accuracy of the analytical model, since $\gamma$ was the only undetermined parameter. This result is sensitive to the power distribution created by the ring heater that is not perfectly axis-symmetric as in our ideal representation. It means that our model slightly overestimates the amplitude of $S$: it might be lowered by the contribution of the ring heater power to some non axis-symmetric high order modes of aberrations. The linearization central to the temperature field model carries through to the aberration model and the spherical component $S$, meaning we expect that $S$ should be proportional to $P$. Looking at the eight transients, there appears a close proportionality between the peaks of $S$ and the power step $P$ applied; the magnitude of the power steps ranges from 2.47 W to 9.84 W, but the values of $\gamma$ agree to within around 10\%, consistent with the expected linearity of the test mass temperature field. Since ETMX and ETMY and the X and Y ring heaters are identical, they should have similar absorption parameters.

\section{Conclusion}
We have derived an analytical model for the temperature field evolution of the test mass due to heating by the ring heater component and determined the resulting time dependent aberration for the general case of a test mass and ring heater. Applying the model to the particular configuration in Advanced LIGO, we have found that HWS measurements of the aberration are in reasonable agreement with the model, and using HWS data we have calculated an estimate for the absorption efficiency of ring heater power of $76 \pm 8 \%$ and $74 \pm 7 \%$ in the Advanced LIGO Y and X end test masses respectively. Combining this model with the model for main laser heating from Hello and Vinet and the HWS measurements provides an understanding of the thermal state of the test masses. Incorporating these models and data into a state estimation filter will allow us to determine the appropriate actuation through the ring heater to best compensate for the thermal lens caused by heating from the main laser beam at any time. These results contribute to the effort to achieve design sensitivity in advanced gravitational-wave detectors by minimizing aberrations from thermal lensing in the test masses.

\section*{Funding Information}
National Science Foundation (NSF) (126289, 1205882).

\section*{Ackowledgement}
The authors thank LIGO Laboratories and especially the LIGO Livingston Observatory staff for additional support. The authors warmly thank Brian O'Reilly and Shivaraj Kandhasamy for the useful discussions and Gabriela Gonzalez for her encouragement.


\end{document}